\documentstyle[preprint,aps,pre,epsfig]{revtex}
\tightenlines

\begin{document}
\title{ A model solid-state structural transformation\,: 
Tetragonal to Orthorhombic}

\author{Madan Rao$^1$ and Surajit Sengupta$^2$}

\address{
$^1$ Raman Research Institute, C. V. Raman Avenue, Sadashivanagar, 
Bangalore 560080, India \\
$^2$ Satyendra Nath Bose National Center for Basic Sciences, 
Block J D, Sector III, Salt Lake, Kolkata 700098, India 
}

\date{\today}
\maketitle

\begin{abstract}

We study equilibrium properties of a system of particles in two
dimensions, interacting with pair and three body potentials, which
undergoes a structural transition from a square to a rhombic lattice and
thus constitutes a simple model for a generic tetragonal to orthorhombic
transition. We aim at an intermediate level of description lying
in-between that of coarse grained elastic strain hamiltonians and
microscopic ab~-initio approaches. Macroscopic thermodynamic properties
and phase diagram at zero and finite temperatures as a function of the
density and the relative strengths of the pair and three body energies are
obtained using lattice sums, an approximate `cell-model' theory and
molecular dynamics simulations in the NVT ensemble. We propose that this
model solid can be used as a test bed for studies of statics and dynamics
of structural transitions.

\end{abstract}


\section{Introduction}

In spite of its fundamental and technological interest, there is as yet no
general theoretical framework to predict the final microstructure of a solid following
changes in temperature or stress across a
structural transition\cite{aron}. This is in part because most experimental
studies have focussed on technologically important solids which are
far-from-idealized\,; 
thus it has been difficult to isolate generic principles amidst the
volume of empirical data\cite{mhb,expts}.
Furthermore, the spatial and
temporal resolution of {\it in-situ} experimental probes is limited\,; this makes
it difficult to follow microstructural changes at short length and time scales.
We believe progress can only be made if we (a) identify
simple `model' systems which would serve as an arena for detailed studies on the
dynamics of structural transitions in solids and (b) develop `probes' to study
the dynamics and morphology changes at short scales\,; in other words to {\it
follow the motion of individual atoms as the transformation proceeds}. 
High speed computational modelling
allows one to make useful contributions to
both (a) and (b). This paper concerns point (a) -- we provide a model system for
a simple structural transition which could serve as a test bed for future studies
of statics and dynamics of structural transitions. In another paper\cite{raosen}
we have discussed point (b) -- we have studied the dynamics of solid state
transformations in this model system by tracking individual particles as the
transformation proceeds.

Our choice of the simple structural transformation, a square to a rhombic lattice
in two dimensions (2-d), is motivated by two important considerations. The first
is its relevance to `real' solid state transformations. The square-to-rhombus
transition may be regarded as a rather accurate
representation\cite{TO-theory,kohn} of the three dimensional tetragonal to
orthorhombic (TO) transition\cite{fe} in an oriented single crystal where the
strain along the third direction (c - axis)  is negligible. This transition is
observed in many technologically important systems\cite{TO-expt}, such as the
high-T$_c$ compound YBa$_2$Cu$_3$O$_7$.

The second motivation is more conceptual and warrants some explanation. Following
a quench across a structural transition, the atoms constituting the solid have to
rearrange themselves, since the parent phase is thermodynamically unstable.
Instead of moving to the new equilibrium configuration, the motion of the
atoms is arrested in a final microstructure which is very far from equilibrium.
The microstructures often display features at length scales ranging from
$1000\AA-100\mu m$, many orders larger than the lattice spacing. A more
appropriate description of the dynamics at these scales is in terms of continuum
degrees of freedom. It is however not clear {\it apriori}, what are the relevant
continuum degrees of freedom, especially in situations where the solid undergoes
large deformations from the parent. In most theoretical studies of this problem,
the only continuum degrees of freedom relevant at these scales have been taken to be the
components of the strain tensor\cite{conti,strain}. However, short length scale
phenomena like atomic rearrangements\cite{strnp,nature} are not captured by these
strain-only theories and may in some cases affect the kinetics of the
transformation. The only unbiased way to determine the complete set of relevant
degrees of freedom in the continuum is to start from a more microscopic
description and arrive at a continuum description using a coarse-graining
method\cite{econst}. Such a coarse-graining program is more easily set up in the
crystallographically simpler square-to-rhombus transition.

How microscopically detailed should our microscopic model solid be ? An ab-initio
or semi empirical description\cite{marsimul} which includes electronic degrees of
freedom tailor-made for a real system such as YBa$_2$Cu$_3$O$_7$ suffers from 3
drawbacks --- (i) it compromises the need for generality, (ii) it is
computationally expensive and (iii) it is difficult to extrapolate to the
continuum. We therefore model an {\em effective} Hamiltonian accurate over
distances smaller than the bulk elastic correlation length but larger than the
typical atomic spacing. This effective Hamiltonian, coarse-grained over the faster
electronic degrees of freedom, will in general have pair and many-body
interactions.

Our model system therefore consists of a set of N `particles' confined in a 2-d box
of volume V at a fixed temperature T. The particles interact with each other via
specific two and three -body\cite{poten} interactions. While the 2-body
interaction stabilizes a rhombic lattice (triangular for {\em isotropic} two-body
potentials), the 3-body interactions have been {\em constructed} to favour a
square lattice. Unlike in a real solid, the square to triangular transition in
our model solid may be driven by {\it independently} (i) increasing the density,
(ii) decreasing the temperature or (iii) decreasing the relative magnitude of the
three-body potential. We confine ourselves here to the equilibrium aspects of
this transition and postpone a detailed study of the nucleation and coarsening
dynamics to a subsequent publication\cite{raosen}.

In the next section we introduce our model, describing the pair and three -body
potentials. In Section III we present the static lattice (zero temperature)
results for the energy, stress and the elastic constants and exhibit the zero
temperature phase diagram. In Section IV we discuss the effect of finite
temperatures. We present results both from a ``cell model''
approximation\cite{celmod,freevol} and from molecular dynamics
simulations\cite{simul} in the NVT ensemble. In Section V we discuss a simple
generalization of the model to include a molecular solid with a complex basis
motif.  Section VI presents a summary and conclusion of this work.

\section{The model potential}
In this section we motivate the form of the {\it effective} hamiltonian,
coarse-grained over the faster electronic degrees of freedom. Instead of starting
out {\it ab-initio}, we shall assume that the solid can be described by a general
functional of the densities of particles. A simple form of the density
functional, proposed by Ramakrishnan and Yussouff (RY)\cite{RY}, views
the solid as an extremely inhomogeneous
liquid with a non-uniform, periodic, coarse-grained density $\rho({\bf r})$. 
The advantage of this approach is that it allows us to make accurate statements
at any finite temperature $k_B T=\beta^{-1}$. Following RY, we may write
the Helmholtz free energy $F_s$ per unit volume $V$ of the solid as, 
\begin{eqnarray}
\frac{\beta F_s}{V} & = & \frac{1}{v} \int_v\,d^3\,r\,\Bigg[ \rho({\bf r})\,\ln\,(\rho({\bf r})/\rho_s) -\rho_s \Bigg] \nonumber \\ 
& & - \frac{1}{2}\,\sum_{\bf G}\rho_{\bf G}\rho_{\bf -G}C^{(2)}({\bf G})
- \frac{1}{3}\,\sum_{\bf G_1,G_2} \nonumber \\
& & \rho_{{\bf G}_1}\rho_{{\bf G}_2}\rho_{-{\bf G}_1-{\bf G}_2}\,C^{(3)}({{\bf
G}_1,{\bf G}_2,-{\bf G}_1-{\bf G}_2}) + \cdots 
\label{dft}
\end{eqnarray}

\noindent
The leading term gives the ideal gas contribution (the integral is over the
Wigner-Seitz unit cell of volume
$v$), while the subsequent terms arise from the
interactions between density waves $\rho_{\bf G}$ with wave-vector
${\bf G}$ belonging to the set $\{ {\bf G} \}$ of reciprocal
lattice
vectors (RLV) of the crystalline solid. Note that we have defined the Fourier
transform as,
\begin{equation}
\rho_{\bf G} = \frac{1}{v}\int_v\, d^3\,r\, e^{i {\bf G \cdot r}}\,\rho({\bf
r})\,.
\label{fourier}
\end{equation}

\noindent
The interaction terms involve the Fourier transform of the $n^{th}$-order 
direct correlation
functions $C^{(n)}({\bf G}_1,{\bf G}_2,\cdots,{\bf G}_{n-1})$.
These functions may either be evaluated from a liquid state theory\cite{hm} or
deduced from
scattering experiments (for instance, $C^{(2)}(q) = 1 - \rho_{q=0}/S(q)$, where
$S(q)$ is
the 
structure 
factor of the liquid at wavenumber $q$)\,; however evaluation (or
measurement) of direct
correlation functions beyond the second order is extremely 
difficult\cite{3body}. 
Approximate ways of incorporating the 
effects of these higher order correlations have been used with varying degrees of 
success\cite{odft}, though often a simple weak coupling
(mean-field) approximation, 
$C^{(n)} = V^{(n)}/k_B T$, 
where $V^{(n)}$ is the $n-$ body potential, works remarkably well\cite{mfdft}. 

\begin{figure}[]
\begin{picture}(30,100)
\put(10,0) {\epsfig{figure=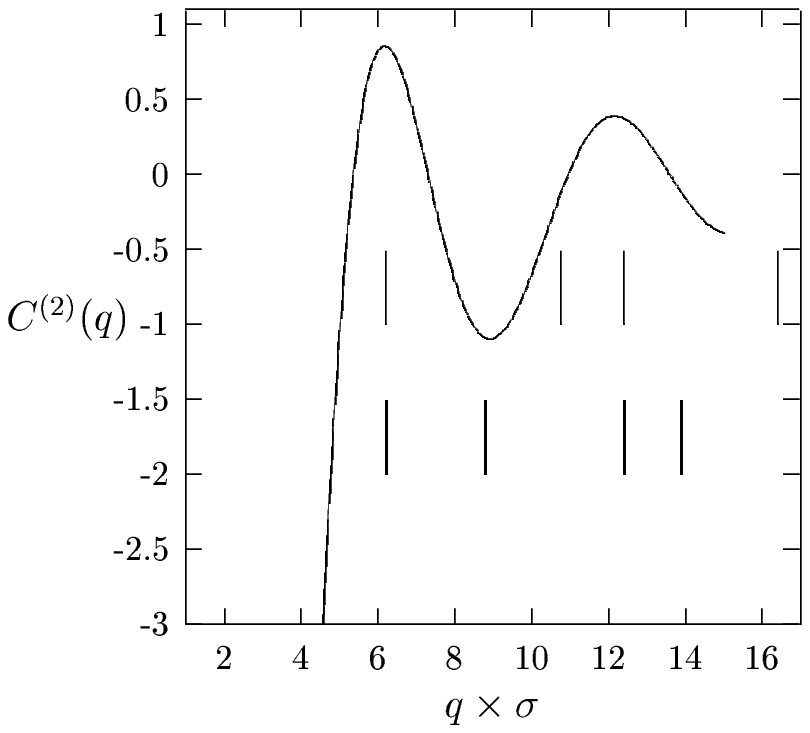,width=90mm,height=90mm}}
\end{picture}
\vskip 2 cm
\label{cofq}
\caption[]
{
Plot of the second order direct correlation function $C^{(2)}(q)$ vs. wavenumber $q$ for a supercooled  hard disk liquid in two-dimensions. The lines mark the
lengths of the reciprocal lattice vectors for the triangular (top) and the
square (bottom) lattices scaled so that the smallest RLV corresponds to the
first peak in $C^{(2)}(q)$.}
\end{figure}

While the ideal gas term always prefers a uniform liquid, the sign of the
interaction term decides whether the  interactions 
stabilize (destabilize) density waves with wave-vectors ${\bf G}\neq0$.
As an
example, let us consider first 
the effect of only the second order terms in Eq.\,\ref{dft}. The coefficients,
$C^{(2)}(|{\bf G}|)$, measure the stability of a density wave with wavenumber
$|{\bf G}|$. In Fig.\,\ref{cofq} we have plotted $C^{(2)}(q)$ against 
$q$ for a slightly supercooled hard disk liquid\cite{rosehd} in two-dimensions.
This function is oscillatory and has a primary peak at roughly the wave number 
corresponding to the magnitude of the smallest RLV of
the thermodynamically stable solid. Choosing the lattice parameter such that 
the smallest RLV coincides with the first peak of $C^{(2)}(q)$, we have 
plotted the positions of the RLVs for the triangular (top) and square (bottom)
lattices in Fig.\,\ref{cofq}. Closed packed lattices have RLVs which are, on
an average, more widely separated than those of an open lattice. As a 
consequence, open lattices often have RLVs lying in the region of the first 
{\it minimum} of $C^{(2)}(q)$ which is negative, thereby contributing to a
destabilisation of the lattice. From 
Fig.\,\ref{cofq} we observe that this is indeed the case for the square lattice
in two-dimensions\cite{bccfcc}. Density waves corresponding to the second RLV 
of the square lattice with Miller index $\{11\}$ are not favoured, making
the square lattice unstable in two-dimensions. 

Note that the discussion above is the finite $T$  generalisation of the zero temperature
result that for isotropic, purely repulsive {\em pair} potentials in two
dimensions, one can only stabilize the close-packed triangular lattice\cite{PCP}.
For instance, a static calculation\cite{born} of the $T=0$ elastic moduli of the
square lattice reveals a shear instability which spontaneously distorts the
lattice till it regains a close-packed structure.

In order to stabilize the square lattice one needs to go beyond the second
order contribution and consider the effect of three-body correlations. As 
is clear from Eq.\,\ref{dft}, a positive (and large enough !) contribution from  
$C^{(3)}({\bf G}_1,{\bf G}_2,{-{\bf G}_1-{\bf G}_2})\rho({\bf G}_1)\rho({\bf
G}_2)\rho({-{\bf G}_1-{\bf G}_2})$, 
where any one (or two) of the wave-vectors equals $\{11\} $, can  
compensate for the destabilizing effect of
the second order correlator\cite{bccfcc}. 
There are many choices for the wave-vectors
${\bf G}$, but the simplest combination is ${\bf G}_1 = {\bf G}_2 = \{10\}$\,\,
(so that $\{10\} + \{01\} = \{11\}$). 
A straightforward way to ensure that this combination of density waves 
is stabilized
is to stabilize the {\em real space} triangle involving the direct 
lattice vectors $(10)$, $(01)$ and $(11)$ (and those related to them 
by symmetry). Within a simple minded mean-field approximation this may be 
accomplished, as shown below, by choosing an appropriate three-body potential 
which favours $0^\circ, 45^\circ$ and $90^\circ$ bonds.
Higher- order interactions involving four or more particles,
though present in principle, are not necessary for our purpose. 

Our model system, constructed from this level of coarse-graining, 
therefore consists of `point-particles' interacting
with effective pair
and three -body potentials.  
The interaction energy $E$ of the system is given by, 
\begin{equation}
E = \frac{1}{2}\sum_{i \neq j} \Psi_2({\bf r}_{ij}) + 
\frac{1}{6}\sum_{i \neq j \neq k} \Psi_3(
{\bf r}_{ij},{\bf r}_{jk},{\bf r}_{ki})\,.
\label{nrg}
\end{equation}
For the pair potential we take,
\begin{equation}
\Psi_2(|{\bf r}_{ij}|) = V_2 \big(\frac{\sigma}{|{\bf r}_{ij}|}\big)^{12} \, ,
\end{equation}

\noindent
which is purely repulsive and therefore the system has to be confined
with a uniform hydrostatic pressure (see below). A 
purely repulsive system simplifies our analysis since there is one fewer 
length scale and one fewer non~-solid phase. Without the 3-body
potential, our system is characterized by only one parameter instead of the
two (temperature and density). On including the 3-body potential, 
we lose this simplification, but the variation of
thermodynamic properties with density is 
still weak. 
Without loss of generality we can take $V_2$ and 
$\sigma$ to be our units for energy and distance respectively.

The 3-body potential is 
\begin{eqnarray}
\Psi_3({\bf r}_{ij},{\bf r}_{jk},{\bf r}_{ki}) & = & V_3\,\,\Big[\,\,f_{ij}\sin^2(4\Theta_i)f_{ik} \nonumber \\ 
&   & \,\,\,\,\,\,\,\,+ \,f_{ij}\sin^2(4\Theta_j)f_{jk} \nonumber \\
&   & \,\,\,\,\,\,\,\,+ \,f_{jk}\sin^2(4\Theta_k)f_{ki}\,\,\,\Big] 
\label{threebody}
\end{eqnarray}

\noindent
where the function, 
\begin{eqnarray}
f_{ij} \equiv f( r_{ij}) & = & (r_{ij} - r_0)^2 \,\,\,\,\,\,\,\, { r_{ij}\,\,<\, r_0} \nonumber \\
& = & 0\,\,\,\,\,\,\,\,\,\,\,\,\,\,\,\,\,\,\,\,\,\,\,\,\,\,\,\,\,{\rm otherwise}
\end{eqnarray}

\noindent
and we have used the notation $ r_{ij} \equiv |{\bf r}_{ij}|$. The angles are 
as defined in Fig.\,2. The function $f_{ij}$ provides a cutoff for the 
3-body potential\,; as long as $f_{ij}$ is short ranged, the actual form of this 
function does not affect the qualitative results. 

It may appear that a three-body potential requires a large investment in terms
of computer times. This apprehension is fortunately unfounded. The form of this 
potential ensures 
that three-body energies can be calculated\cite{poten} extremely efficiently, 
requiring a computational effort not exceeding that for the pair part. This 
is discussed in the Appendix.

\section{Zero temperature results}

\begin{figure}[]
\begin{picture}(30,70)
\put(0,0) {\epsfig{figure=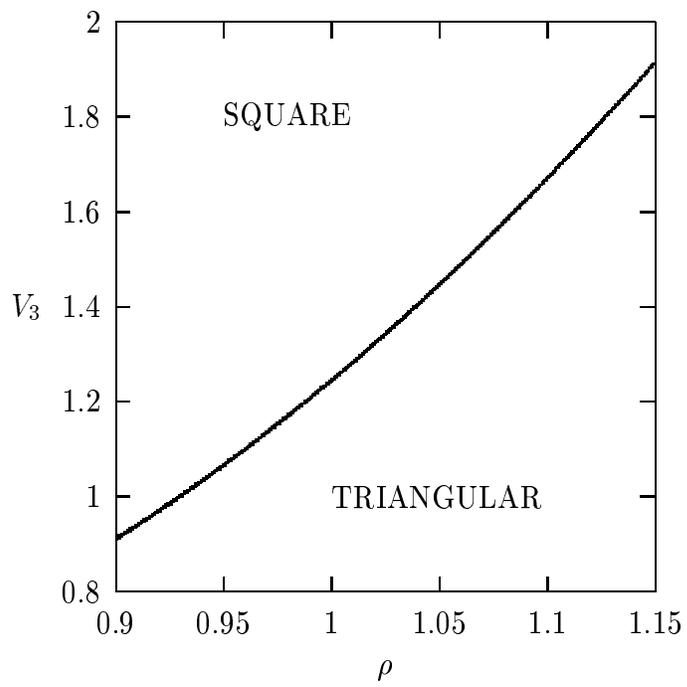,width=90mm,height=90mm}}
\end{picture}
\vskip 2 cm
\caption[]
{\label{t0phase}Zero temperature phase diagram in the $V_3 - \rho$ plane. The
regions
where the square and the triangular phases are stable are labelled.}
\end{figure}

At zero temperature, the equilibrium configuration is a solid which minimises the
potential $E$. Since we work in the constant NVT (and shape) ensemble, the density
$\rho = N/V$ is a constant.
Assuming that the only minima of $E$ correspond to the triangular
or square phases, we have numerically deduced the $T=0$ 
phase diagram in the $\rho - V_3$ plane (Fig.\,\ref{t0phase}).
Later in this section we show, within a restricted variational calculation, that
these are the only minimisers of $E$.
As we see from Fig.\ \ref{t0phase}, the triangular lattice is the lowest
energy 
phase at high densities and small values of $V_3$, {\it i.e.,} wherever the pair
interaction dominates over the three-body part. Across the boundary there
is a strong first-order transition with a discontinuous change in the slope
$\left(\partial E/\partial \rho\right)_{V_3}$.

To deduce the nature of the order parameter distinguishing the square from the
triangular phase, we look at how a rhombic lattice may be obtained from a square.
Such an analysis makes contact with continuum elasticity in a natural way.

At zero temperature, a continuous family of perfect rhombic lattices 
(labelled by position vectors ${\bf R}^T$)
can be obtained from
the perfect square lattice (labelled by ${\bf R}^0$)
by the transformation,
$
{\bf R}^T = ({\sf I}+{\sf T}){\bf R}^0
$
where the transformation matrix {\sf T} is,
\begin{equation}
{\sf T }  =  \Bigg( \begin{array}{lr}
\epsilon_1/2 + \epsilon_2/2 & \epsilon_3  \\ 
\epsilon_3  & \epsilon_1/2 - \epsilon_2/2  
\end{array}
\Bigg).
\label{matrix}
\end{equation}

\begin{figure}[]
\begin{picture}(30,100)
\put(0,0) {\epsfig{figure=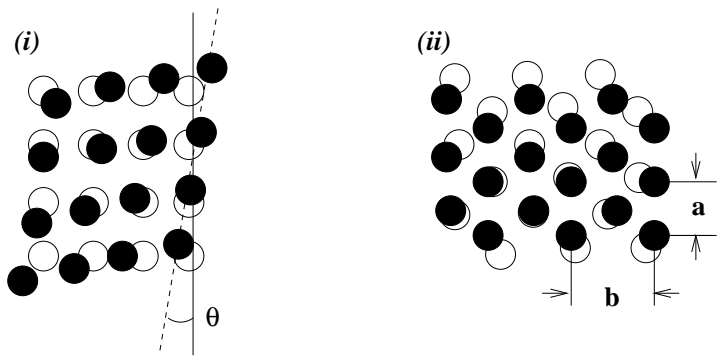,width=100mm,height=50mm}}
\end{picture}
\vskip 2 cm
\caption[]
{\label{trian} Two equivalent ways of obtaining a triangular lattice (filled
circles) from
a square (i) and (ii). One can either (i) shear the original
lattice by an angle $\theta$ or (ii) rotate the original
lattice by $45^\circ$ and then stretch it along one of the axes and
compress it along the other so that $b/a > 1$.
For a square lattice $\theta = 0, b/a = 1$ and
$\theta = 15^\circ, b/a = \sqrt 3$ for the ideal triangular lattice. In terms ofthe shear strain $\epsilon_3$ the corresponding numbers are $0$ and $.27812$.
}
\end{figure}

\noindent
The parameters $\epsilon_{\alpha}$ ($\alpha=1,2,3$) are related to the
components of a strain tensor by the following construction.  
We choose to measure all distortions and energies with respect to the 
undistorted {\em square} phase --- our reference state. The microscopic
displacements ${\bf u_{
R^0}} = {\bf R}^T - {\bf R}^0$ are therefore defined at every ${\bf R^0}$ on the
reference lattice, {\it i.e.,} we use 
{\em Lagrangian}\cite{PCP,landl} coordinates. 
The full nonlinear Lagrangian strain tensor\cite{landl} $\epsilon_{ij}$ is,
\begin{equation}
\epsilon_{ij} = \frac{1}{2} \left( \frac{\partial u_i}{\partial r_j} +
\frac{\partial u_j}{\partial r_i}+\frac{\partial u_k}{\partial r_i}
\frac{\partial u_k}{\partial r_j} \right)\,,
\label{defstrain}
\end{equation}
where the indices $i,j$ go over $x$ and $y$.                         
The parameters $\epsilon_{\alpha}$ in Eq.\,\ref{matrix}  
represent the combinations 
$\epsilon_{xx}+\epsilon_{yy}$, 
$\epsilon_{xx}-\epsilon_{yy}$ and $\epsilon_{xy}$ 
respectively, which reduce to the usual volumetric, deviatoric and shear strains once 
nonlinearities are neglected. Note that one may start with a prescribed square and end
with a 
final triangular lattice in more than one way ---
the transformation parameters are not unique.
For instance, for a given orientation of the parent square 
lattice shown in Fig.\,\ref{trian}(i) one obtains a rhombic lattice using
$\epsilon_2 = 0$
and $\tan \theta = \epsilon_3/(1+\epsilon_1)$. Equivalently
(Fig.\,\ref{trian}(ii)) the 
square lattice (first rotated by 
$45^\circ$) can be transformed to a centered rectangular lattice with 
$\epsilon_3 = 0$ and 
$b/a = (1+\epsilon_2+\epsilon_1/2)/(1-\epsilon_2+\epsilon_1/2)$. The two transformations 
are completely equivalent. 

One of the ofshoots of this nonuniqueness, is that
any rhombus obtained as a uniform deformation of a perfect square,
can be represented by two independent parameters 
$\epsilon_{1}$ and $\epsilon_{3}$. In addition, since the density
$\rho$ is a constant in our NVT ensemble, $\epsilon_1$
is related to $\epsilon_3$; $\epsilon_1 =
\sqrt{1+\epsilon_3^2}-1 
\approx  \epsilon_3^2/2$. Thus
all rhombic lattices considered by us can be labelled by a
single parameter $\epsilon_3 = \epsilon$ (which by definition is $0$ for the
perfect square lattice). This makes the shear strain $\epsilon$ a good order
parameter which
distinguishes the square from the triangular lattices.

\begin{figure}[]
\begin{picture}(30,100)
\put(0,0) {\epsfig{figure=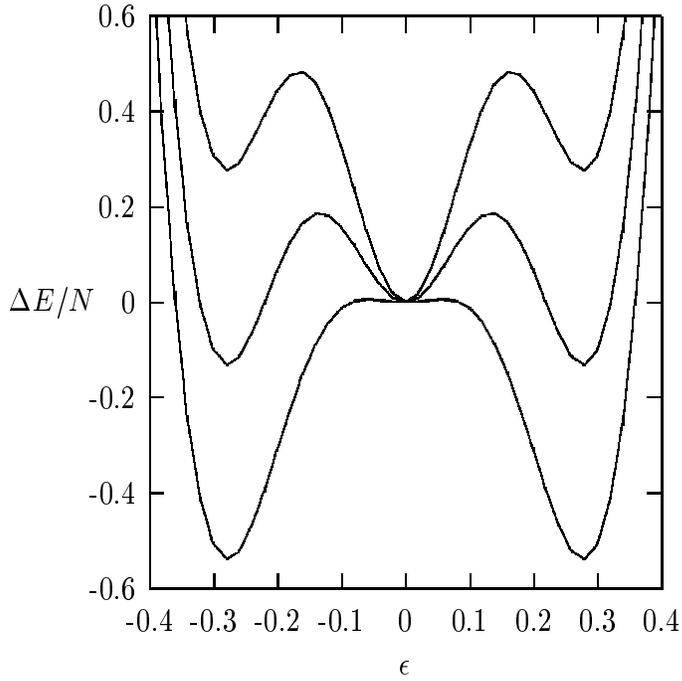,width=90mm,height=90mm}}
\end{picture}
\vskip 2 cm
\caption[]
{\label{energy} Energy difference per particle ($\Delta E/N$) between the square 
and rhombic lattices as a function of the strain order
parameter $\epsilon$ (see text).  Of the three minima shown,
the one at $\epsilon = 0$ corresponds to the square phase and the two
degenerate minima at $\epsilon = \pm \epsilon_{0}$ corresponds to two different
orientations of the triangular phase.
The curves are for
$V_3 = 2.0$ (top), $1.5$ and $1.0$. Note the first order transition
from the square to the triangular phase
as $V_3$ is reduced.
}
\end{figure}

We may now calculate the energy of $T=0$ configurations as a function of
the order parameter $\epsilon$.
A calculation of the energy $E$ and its derivatives (elastic moduli) for a
given lattice involves computation of lattice 
sums. We start with a finite square lattice containing $10\times10$
sites and allow the
transformation
${\sf T}$ to produce a continuous sequence of rhombic lattices
labelled by $\epsilon$. 
We have checked for convergence of the lattice sums by increasing the
size of the lattice and observing the consequent change in the numerical values. 
In Fig.\,\ref{energy} we have plotted the energy per particle $E/N$
as a function of 
the parameter $\epsilon$ for various values of $V_3$ (keeping $\rho$ fixed). Note
that for large values of
$V_3$ there is only one minimum at $\epsilon = 0$ so that the square lattice is the only 
stable phase. For smaller values
of $V_3$ two additional degenerate minima appear at $\epsilon = \pm \epsilon_{0} 
= \pm 0.27812$ which correspond to the triangular lattice. The transition is
first-order with the order parameter jumping discontinuously,
$\vert \Delta \epsilon \vert = \epsilon_{0}$. Note that in this model the change in the
shear
strain
across the transition is fixed\,; we shall return to this point in
Sect. V where we propose a variant of this model in which the jump in the shear strain
across the structural transition can be made arbitrarily small. 
As expected, the square and 
triangular lattices are the only minimisers of $E$ within this parametrisation 
scheme.

\begin{figure}[]
\begin{picture}(30,100)
\put(0,0) {\epsfig{figure=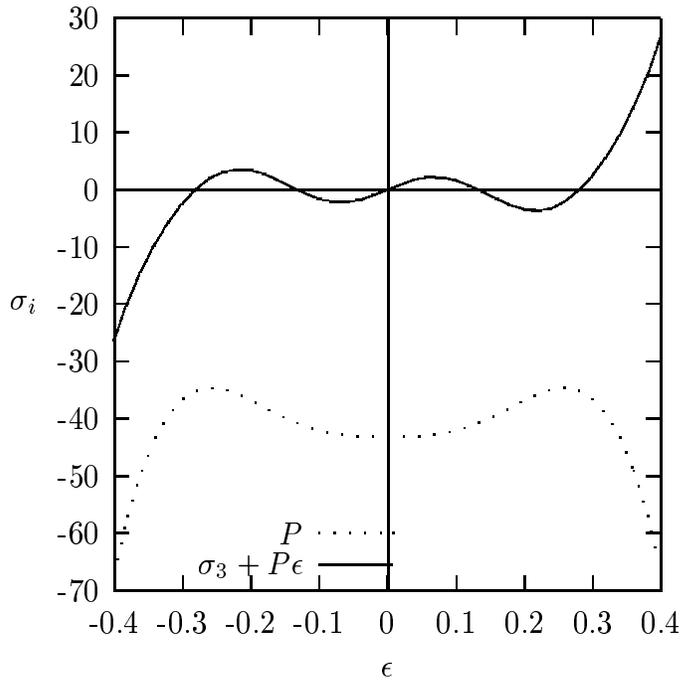,width=90mm,height=90mm}}
\end{picture}
\vskip 2 cm
\caption[]
{\label{stress} The pressure $P = \sigma_{1}$ and the ``effective'' shear stress$\partial E / \partial \epsilon = \sigma_{3} + P \epsilon$ as a function of
$\epsilon$ for $\rho = 1.1$ and $V_3 = 1.5$.}
\end{figure}

\begin{figure}[]
\begin{picture}(30,100)
\put(0,0) {\epsfig{figure=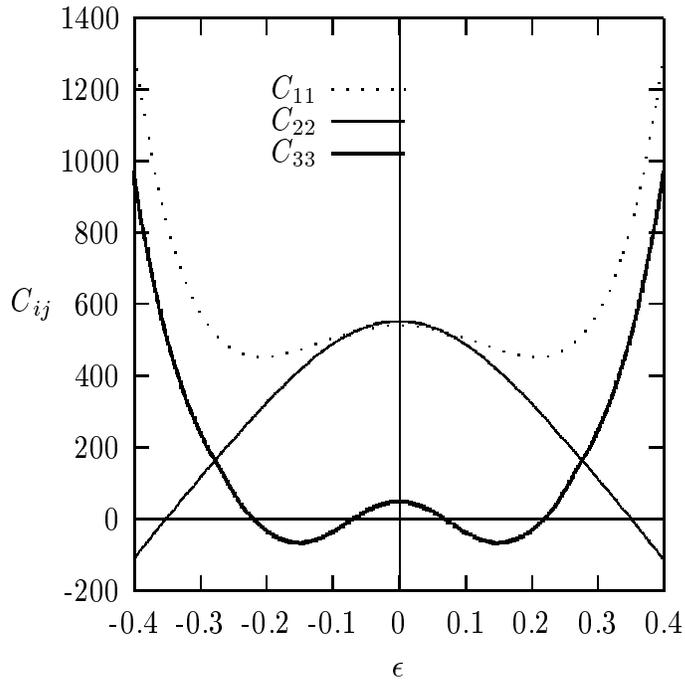,width=90mm,height=90mm}}
\end{picture}
\vskip 2 cm
\caption[]
{\label{moduli} The $2^{nd}$ order elastic moduli $C_{11}$ (bulk), $C_{22}$ and
$C_{33}+P$
(shear) as a function of $\epsilon$. Note that for the triangular
lattice $C_{22} = C_{33}+P$ as required by symmetry. The density $\rho = 1.1$
and $V_3 = 1.5$}
\end{figure}

To make contact with elasticity theory we may compute 
stresses and elastic moduli, obtained by evaluating appropriate 
derivatives of the energy keeping $T,N,V$ constant ---
$ \partial E/\partial \epsilon_{\alpha} = \sigma_{\alpha}$ (Fig.\,\ref{stress}),
and 
$\partial^2 E/\partial \epsilon_{\alpha} \partial \epsilon_{\beta} = C_{\alpha
\beta}$ (Fig.\,\ref{moduli}). 
Note that our system is always under a hydrostatic pressure $P = \sigma_1$\,; 
the constant density constraint implies that for $\epsilon \neq 0$ there is 
an applied shear stress $\sigma_3 = P(\epsilon) \epsilon$ (Fig.\,\ref{stress}). This
implies that the
slope of the
shear stress {\it vs}\,\,$\epsilon_3$ curve is not the shear modulus $C_{33}$
(defined for 
zero external stress) but $C_{33}+P$ (Fig. \ref{moduli}). 

At this stage, we find it useful to point out that the 
results of
Figs. \ref{energy}-\ref{moduli} can be rationalized using a systematic power
series expansion of
the energy in terms of $\epsilon_{\alpha}$. Although such expansions are quite 
common in the literature\cite{TO-theory,strain}, our results show that fourth
order terms in 
$\epsilon_{\alpha}$, especially cross couplings of the 
form $\epsilon_1^2\epsilon_3^2$ and $\epsilon_2^2\epsilon_3^2$, together
with coupling to the external hydrostatic pressure $P$ need to 
be included in order to reproduce the $T=0$ results accurately. The
coefficients 
of all these terms are however not independent. For instance, relationships
like,
\begin{equation}
\frac{\partial^2 F}{\partial \epsilon_2^2} \, \bigg|_{\epsilon_0} = 
\frac{\partial^2 F}{\partial \epsilon_3^2} \, \bigg|_{\epsilon_0}
\end{equation}
dictated by the geometry of the triangular phase have to be satisfied
for all temperatures and densities.

In this section we have been able to show that our model potential indeed 
produces the square and the triangular lattices as minima of the energy.
The potential parameters maybe tuned, if necessary, to a real 
system by comparing the elastic properties of this model system to experimentally
measured quantities. 
By varying the density or the strength of the three-body potential one  
obtains a zero temperature first order structural transition between a 
square and triangular lattice. What happens to the structural transition at 
non-zero temperatures? We study this question in the next section.

\section{Nonzero temperature results} 

In this section we analyse the phase diagram at $T\neq0$ as a function of $V_3$ or
$\rho$. We
do this by two methods --- an `exact' molecular dynamics (MD) simulation\cite{simul} in the
constant NVT ensemble, using the 2- and 3-body potentials defined earlier, and an
approximate `cell-model'\cite{celmod} based on the deformation parameter $\epsilon$.
The latter leads naturally to an approximate continuum elasticity description at $T\neq0$.
We take up the cell-model analysis first and compare its results with the exact
MD simulation in the next subsection.

\subsection{Cell model approximation: free energies and phase stability}

Imagine being in a region of the zero temperature parameter space $V_3 - \rho$, 
where the square solid is the stable minimum of the energy. 
As the temperature is gradually increased, the contribution of the phonon entropy 
to the (Helmholtz) free energy destabilizes the square lattice. In order
to quantify this effect one needs to go beyond the static lattice 
and consider phonon fluctuations. Although a direct calculation of
the contribution of phonons to the lattice energy 
is straightforward\cite{born}, we choose to use the much simpler, though not necessarily
less accurate `cell-model' approximation. 

Before discussing the cell-model approximation, let us mark its regime of validity.
First, the cell-model approximation neglects contributions from topological defects like
dislocation-antidislocation pairs and thus breaks down near the melting
point\cite{econst}. In two dimensions, there is a further complication, since
fluctuations of the displacement field ${\bf u}$ due to phonons diverge
logarithmically\cite{PCP} with system size. This divergence is however weak and may be
ignored for the system sizes under consideration.

Recall that at $T=0$, the configurations of the perfect rhombi were parametrised
by a single deformation variable $\epsilon$. Is this true at $T\neq0$, when the
lattices are not perfect due to phonon fluctuations ? It turns out that the
constraints of rhombic symmetry and constant density still
allow a parametrization of the $T\neq0$ configurations by a single {\it function}
$\epsilon({\bf x})$, at least when the temperatures are low. Thus the energy
$E$ may be written as a functional of $\epsilon$, which at low temperatures
may be replaced by its mean $\langle \epsilon \rangle$.

Within the cell - model approximation, the partition function of a 
lattice of $N$ particles at temperature $T$ is given by\cite{celmod},
\begin{eqnarray}
Z(\langle \epsilon\rangle, T, N) & = & \big[\Lambda^{-3}\int_{v_\epsilon}d{\bf
r}\,\exp(-\delta \phi_\epsilon({\bf r})/k_B T)\big]^N \,\, \times \nonumber \\ 
 &  & \,\,\,\exp(-E(\langle \epsilon\rangle)/k_BT)
\end{eqnarray}

\noindent
where $\Lambda$ is the thermal wavelength and 
$\delta \phi$ is the change in potential energy of a single particle 
as it moves around within a unit cell of size $v_\epsilon$ in a potential 
well arising from its interaction with all its neighbours. A further harmonic
approximation for $\delta \phi$ leads to the familiar Einstein approximation.
At the other extreme, for the hard disk potential, $\delta \phi = 0$ except 
where overlaps occur and the cell-model approximation becomes identical 
to the {\it free volume}\cite{freevol} theory. The Helmholtz free energy for any rhombic 
lattice labelled by $\langle \epsilon\rangle$ may now be  obtained by using $F = -k_B T
\log Z$. 

\begin{figure}[]
\begin{picture}(30,100)
\put(0,0) {\epsfig{figure=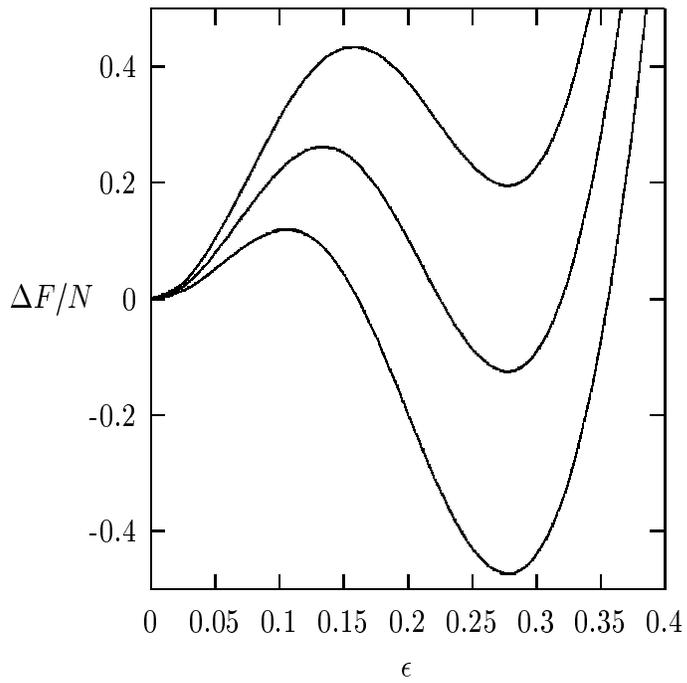,width=90mm,height=90mm}}
\end{picture}
\vskip 2 cm
\caption[]
{\label{cell} The per particle Helmholtz free energy difference ($\Delta
F/N$) between the
square ($\langle \epsilon \rangle = 0$) and rhombic lattices as a function of
$\langle \epsilon
\rangle$ for
various temperatures $T = .1$ (top), $.5$ and $.1$. Note the first order
phase transition from square to triangular with increasing temperature.}
\end{figure}

Evaluation of the Helmholtz free energy (Fig.\ \ref{cell}) allows us to calculate the
$V_3 -T$ phase diagram at any density $\rho$ as also the limits of metastability
of the square lattice.

\subsection{Molecular dynamics simulations}

\begin{figure}[]
\begin{picture}(45,100)
\put(-20,0) {\epsfig{figure=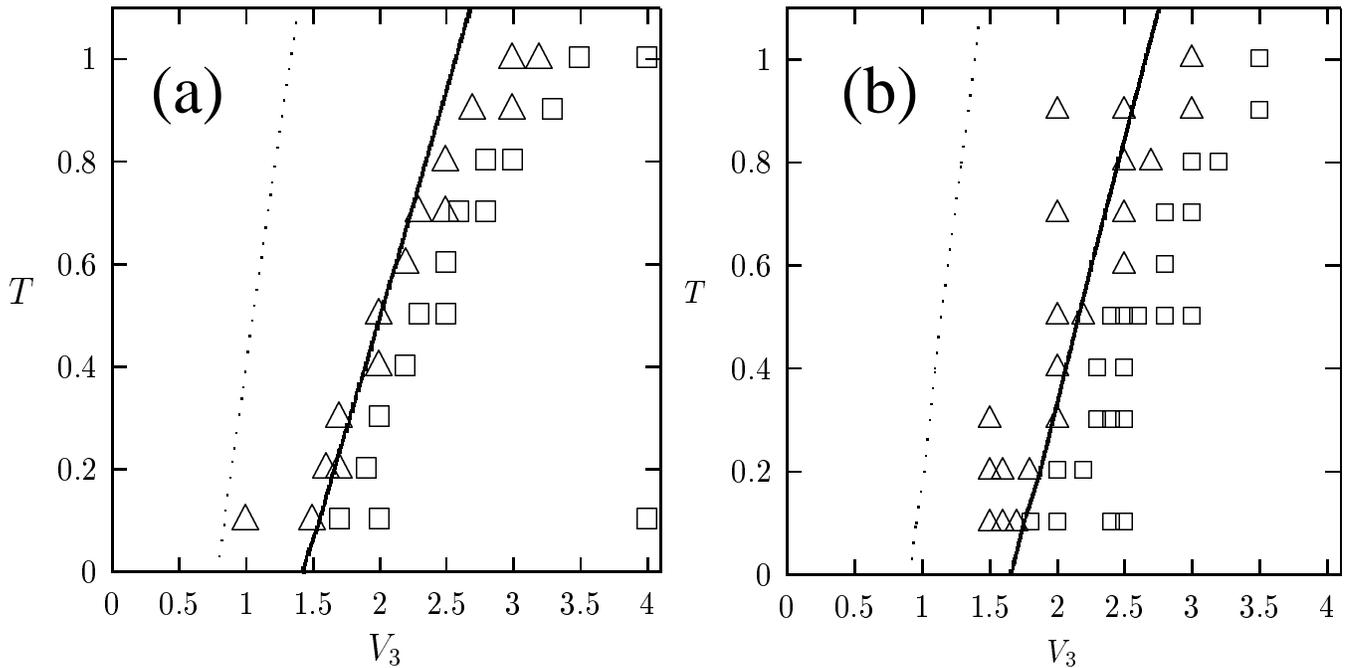,width=180mm,height=90mm}}
\end{picture}
\vskip 2 cm
\caption[]
{\label{md}
Phase diagram in the $T - V_3$ plane for $\rho = 1.05$ (a) and $1.1$ (b).
For large $V_3$ the square phase is stable while the triangular phase is
stable for
smaller values of $V_3$. The points are results from our molecular
dynamics simulations in the $NVT$ ensemble with $2499$ particles. Starting
from an initial ideal square lattice the system was equilibrated for
upto $60000$ steps and the final structure noted ($\Box$ for square and
$\Delta$ for triangular) for various values of
$T$ and $V_3$. The solid line is the phase boundary resulting from the
cell model approximation (see text) and the dashed line is the metastability
limit for the square phase from the same theory.
}
\end{figure}

To obtain accurate results for the phase stability at non-zero
temperatures, we have used a molecular dynamics simulation for our model system. 
We simulate $N = 2499$ particles ($50 \times 50$ unit cells 
with vacancy to improve the kinetics) in the NVT ensemble using a standard
Nos\'e-Hoover thermostat. Starting from an ideal square lattice, we have 
equilibriated systems at various values of $V_3$ and temperature for a fixed 
density for about $50~-100\,\times\,10^3$ molecular dynamics time steps or till 
thermodynamic
quantities like the pressure and energy have stabilized. The final structure 
is then examined and this information used to obtain the phase diagrams shown 
in Fig.\,\ref{md}. We display the phase diagram for two densities $\rho =
1.05$ and $1.1$.
Together with the molecular dynamics results we have also plotted
the results of the cell-model approximation. We observe that for low 
temperatures, the cell-model approximation faithfully reproduces the 
actual phase boundary while  at higher temperatures it begins to deviate.
The cell-model approximation is also used to plot the limit of 
stability of the square phase in the triangular region. 
 
Both the molecular dynamics simulations and the cell model calculations predict 
that the square to triangular transition remains first-order over a wide 
region of parameter space even at non-zero temperatures.  
For larger density, the transition point shifts to higher values of $V_3$. 
This is expected since a high density favours the triangular lattice. Also,
the square phase becomes unstable for lower values of $V_3$ as the density 
is increased. The jump in the order parameter remains fixed at $\vert \Delta
\epsilon \vert = \epsilon_{0}$
all along the transition line. This aspect of our model is  
specifically addressed in the next section, where we show that inclusion 
of an anisotropic {\em pair} interaction allows one to tune the 
order parameter jump all the way to zero. 

We end this section with the following observations.  We have seen that the exact MD
and approximate cell model
give qualitatively similar results. More sophisticated phonon fluctuation calculations
may even produce quantitative aggreement. Figure \ref{cell} suggests that the 
Helmholtz free energy may be expanded in powers of $\epsilon$, just as was noted at
$T=0$. Though we do not explicitly demonstrate here, we may recover 
elasticity theory (including corrections arising from thermal fluctuations) by
constructively coarse-graining as in Ref.\ \cite{econst}.

\section{Generalization to molecular solids}
The model 2-d solid discussed in the preceding sections has the virtue that it
is simple enough to begin a detailed theoretical
study of both the equilibrium and dynamical features of the TO transition across a
range of length and time scales. However if we were to compare the results
of such a study with experiments on realistic systems, we would immediately face a problem. 
Most solids undergoing a TO transition\cite{TO-theory,kohn}, for instance  
YBa$_2$Cu$_3$O$_7$, 
have a complex basis, consisting of many 
atoms per unit cell. These systems generically have much smaller jumps in the 
shear strain at the TO transition compared to the jump computed in the previous
section. To appreciate the {\it quantitative} discrepancy,
recall the discussion following Eq.\ \ref{defstrain}, where we showed that
any perfect 
rhombus obtained as a deformation of a square may be parametrized by either $\epsilon$ or
$b/a$. Defining an 
{\it orthorhombic distortion} as $D\equiv (b - a)/(b + a)$, we find that
$D  = ({\sqrt
{3}}-1)/({\sqrt{3}}+1)\approx 0.27$ for our model square-to-triangle structural transition --
significantly larger than $D= 0.0085$
for the TO transition in YBa$_2$Cu$_3$O$_7$.

Apart from this, there might be a more serious {\it qualitative} mismatch between our model
solid and real systems undergoing a TO transition. Changes in temperature or pressure
(hydrostatic or chemical) may lead to a local structural rearrangement (optical modes) which
would couple to the strain tensor.  The jump in the shear strain across the structural
transition may therefore, unlike in our model solid, vary along the phase boundary, even
going to zero (phonon softening) at a critical point\cite{softy}. 
 
We shall see that we may address both these issues within an anisotropic variant of our
model solid. Our attempt will be to incorporate the complex basis, with many atoms per unit
cell, into an {\it effective} hamiltonian between `point particles'. In the spirit of an
effective hamiltonian, we will coarse-grain the density over a length and time scale
corresponding to the `size' $\xi$ and relaxation time $\tau$ of the basis. Thus we may
define a coarse-grained density as $\rho({\bf r}) = p^{-1} \sum_\mu \rho_\mu({\bf r})$,
where $\mu = 1, \ldots , p$ labels the atomic species making up the basis. This
coarse-grained density profile $\rho({\bf r})$ will have peaks at the centre of mass of each
basis, falling off to zero over a length scale $\xi$ and having a cross section which is
spatially anisotropic. If we assume that this anisotropic cross section has a fixed
shape at a given
temperature and pressure (true when the associated optical branch is much higher than
the acoustic
branches), then we may write the effective hamiltonian as arising from a collection of
`point particles' interacting via an anisotropic potential.
The form of the effective hamiltonian may also be motivated in terms of a density
wave picture
\cite{mangal} in much the same way as in Section II. An anisotropic density
interacts  via
an anisotropic direct correlation function. Within a mean field  
approach this reduces to a  
pair potential which depends not 
only on the distance between the two basis motifs but also on their 
orientation relative to the crystal axes --- orientation fluctuations within the 
motifs being neglected. 

\begin{figure}[]
\begin{picture}(30,100)
\put(0,0) {\epsfig{figure=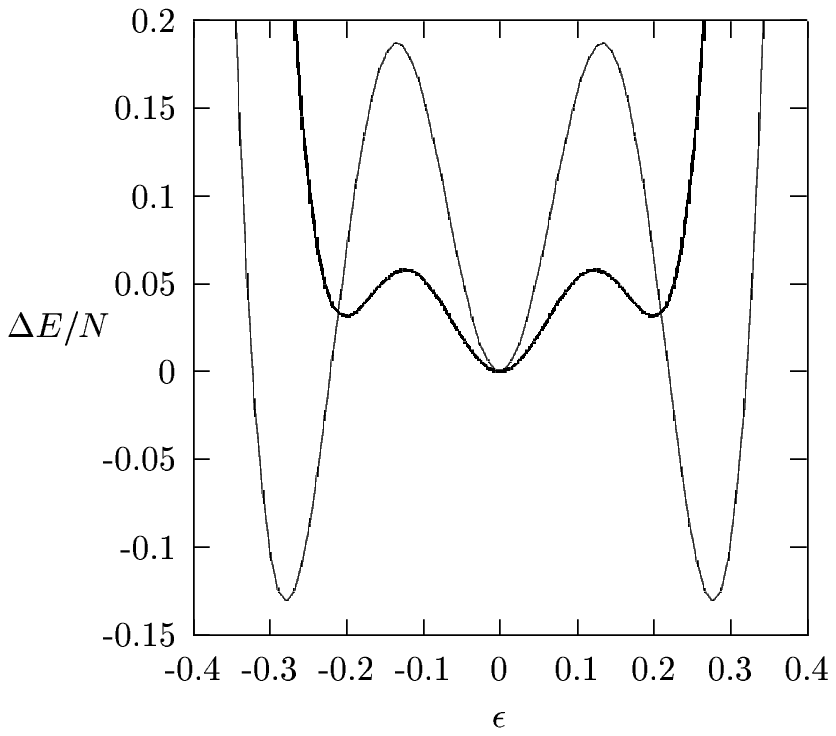,width=120mm,height=120mm}}
\put(45,15) {\epsfig{figure=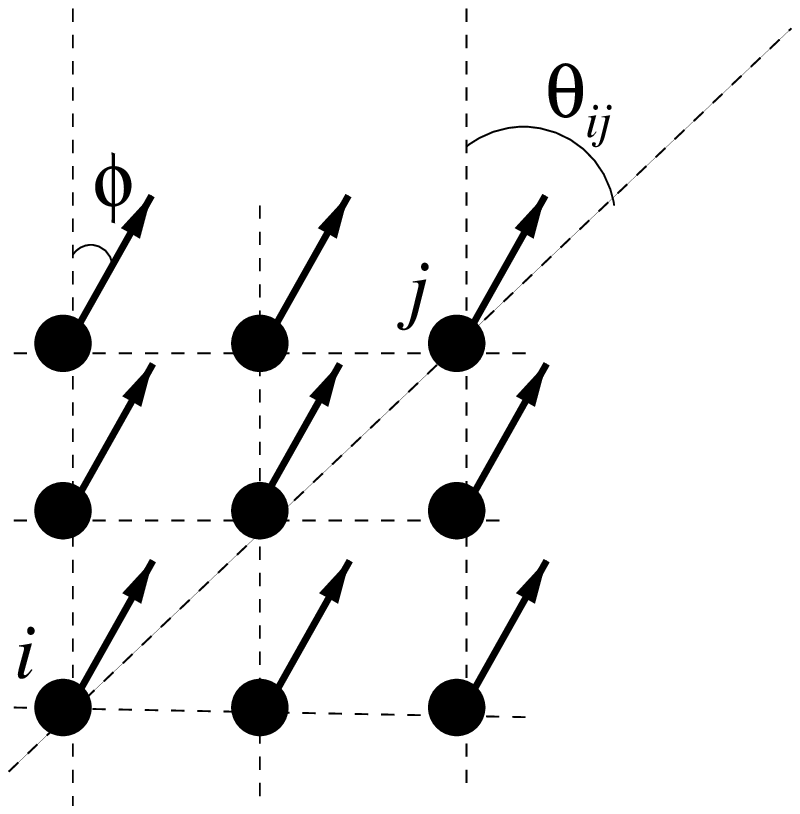,width=50mm,height=50mm}}
\end{picture}
\vskip 2 cm
\caption[]
{\label{molsol}
The energy difference $\Delta E /N$ for two values of the anisotropy
parameter $\alpha = 0$ and $1$  for $V_3 = 1$ and $\rho = 1.1$. Note
that the energy minima for $\epsilon \neq 0$ shifts to lower values
of $\epsilon$ as $\alpha$ increases.
(inset) The meanings of the angles $\theta_{ij}$ and $\phi$ used in the text.
While $\theta_{ij}$ is the angle of the position vector ${\bf r}_{ij}$
between the
molecules $i$ and $j$ measured with respect to the crystal axis $\{0 1\}$ in
the reference square lattice, $\phi$ is the orientation of the basis molecule
measured with respect to the same axis.
}
\end{figure}

For specificity if we assume 2 atoms per basis, we may then arrive at
the following modification of the two-body potential using the arguments outlined
above,
\begin{equation}
\Psi_2(|{\bf r}_{ij}|) = V_2 \big(\frac{\sigma}{|{\bf r}_{ij}|}\big)^{12}\,\times
\left(1 + \alpha \sin^2 4 (\theta_{ij} - \psi)\right) \, ,
\label{eq:modpsi2}
\end{equation}

\noindent
where the anisotropy parameter $\alpha$ has a fixed value at constant $T$ and
$P$. On the other hand, the 3-body potential may be taken to be the same as in Eq.\
\ref{threebody}.
Setting $\gamma = 1 + \alpha$,
we see that $\gamma$ is always positive.
All angles are measured with respect to the $[ 0 1]$ axes of the undistorted
square lattice. The angle $\psi$ represents the orientation of the basis 
and $\theta_{ij} = \sin^{-1} (x_{ij}/\vert{\bf r}_{ij}\vert)$, see
Fig.\,\ref{molsol}. The
total 
energy is a 
function of $\psi$ so  that uniform rotations of the basis with respect
to the 
crystal axes cost energy (optical mode) while simultaneous rotations of the 
basis together with the crystal axes is a symmetry of the hamiltonian. 

Using the modified two-body potential (Eq.\,\ref{eq:modpsi2}), we compute 
the energy as a 
function of $\epsilon$ as in Section III. For a fixed $\alpha$, the total energy
minimized with 
respect to $\psi$ and $\epsilon$ leads to $\psi = 0$
(independent of
$\epsilon$). As before there are 3 minima in $\epsilon$, one at $\epsilon=0$
(corresponding to a square)
and the other two corresponding to rhombi with $\vert \epsilon\vert$
being {\it smaller} than the value for the perfect triangular lattice $\sim .28$
(Fig.\,\ref{molsol}). The jump in the value of the shear strain $\epsilon$ across
the structural transition 
is therefore smaller than that obtained in Section III. Moreover we find that
this jump in
$\epsilon$ goes to zero and the region over which the
square phase is metastable shrinks and disappears as $V_3 \to 0$ thus 
indicating a {\it continuous transition} at a tricritical point. 
One expects, therefore, that for real systems fluctuation effects near the 
T-O transition would be more pronounced. This fact is actually bourne out
by experiments\cite{TO-theory,TO-expt}.

\begin{figure}[]
\begin{picture}(30,70)
\put(-5,0) {\epsfig{figure=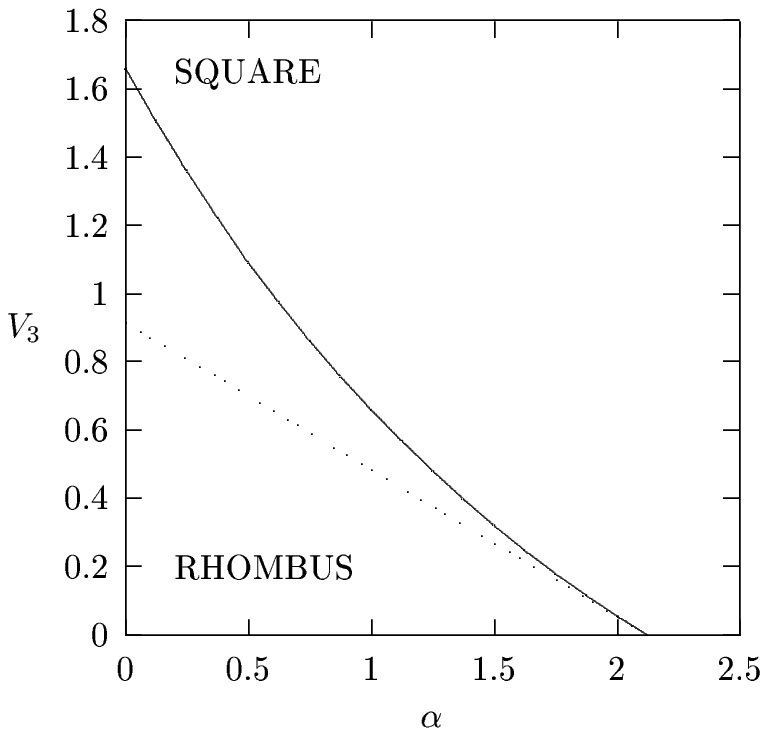,width=80mm,height=80mm}}
\put(26,34) {\epsfig{figure=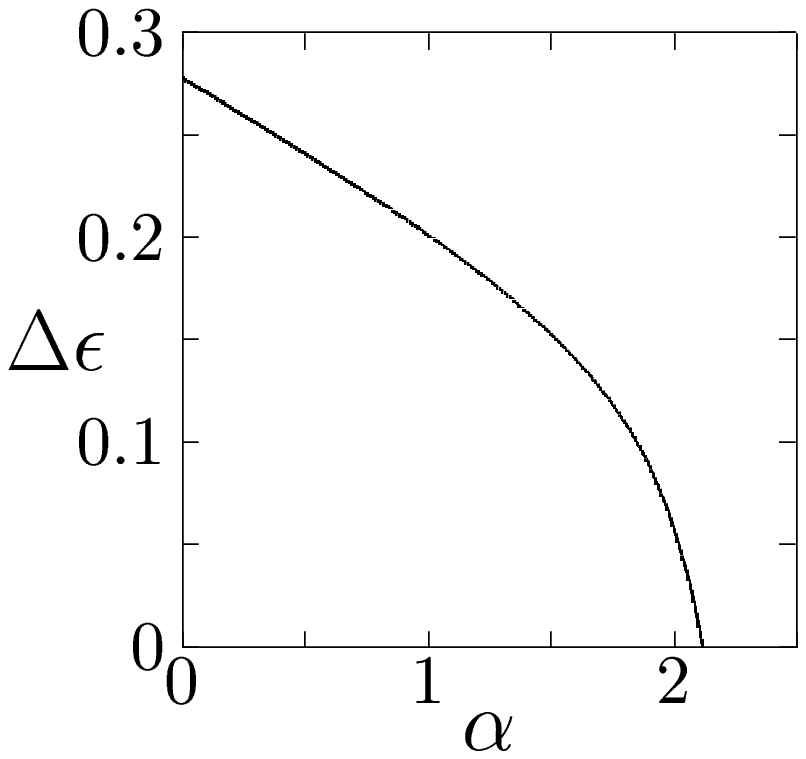,width=40mm,height=40mm}}
\put(78,0) {\epsfig{figure=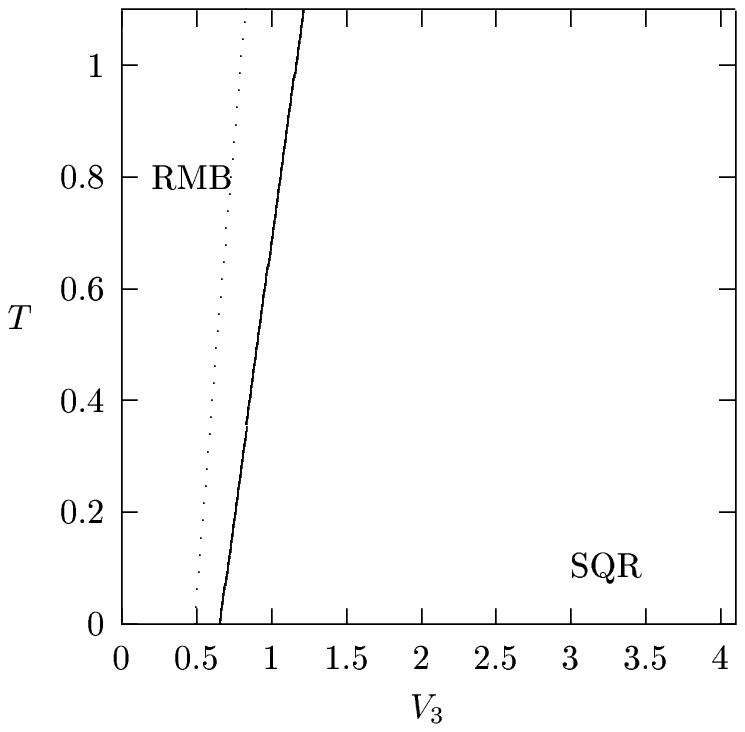,width=80mm,height=80mm}}
\end{picture}
\vskip 2 cm
\caption[]
{\label{phasediag}
(left)The zero temperature phase diagram in the $V_3 - \alpha$ plane for
$\rho = 1.1$. The dashed
line marks the limit of metastability of the square phase. Note that for
$V_3 = 0$ reducing $\alpha$ produces a second order transition with a
tricritical point at $\alpha_{tc} = 2.24$. The inset shows the jump in the
order parameter $\Delta \epsilon$ accross the square~-rhombus phase boundary
as a function of the anisotropy $\alpha$. (right)The phase diagram in the
$V_3 - T$ plane for $\alpha = 1$ and $\rho = 1.1$.
The dashed line marks the limit of metastability of the square phase.
Note that in real systems $\alpha$ depends on $T$ so that, in general, any
quench traverses
a trajectory in the parameter space $T-\alpha-V_3$ and the first-order line
may end in a non-zero temperature tricritical point.
}
\end{figure}

The zero and finite temperature phase diagrams are shown in 
Fig.\,\ref{phasediag}.
The zero temperature phase diagram clearly shows the location of the 
tricritical point where the jump in the order parameter vanishes.
The effect of finite temperatures is addressed easily within a 
cell-model approximation. The calculation may be carried out along the lines 
outlined in the last section. Once again we see that the Helmholtz
free energy can be written as a power series expansion in $\epsilon$.
The results of the calculation are expected to 
be accurate at low temperatures if the anisotropy is not too large. 
For larger anisotropies the effect of (tri)-critical fluctuations may 
alter the results of our simple mean-field estimates. Our result shows 
that the square lattice now becomes stable over a much larger range of 
$V_3$ than in the isotropic case. The region of metastability of the square
lattice however decreases and the first-order transition is weakened.

\section{Summary and conclusion} 
In this paper we have described a model system which is designed to undergo 
a square to rhombus transition in two dimensions. We believe that our 
study will be useful in two ways. On the one hand, it may be used as a 
simple simulational model for the T-O transition in real materials which
often consists
of a large number of individual atomic species making it difficult to study 
using ab-initio methods. For this purpose, the parameters $V_3$ and $\alpha$ 
have to be ``fitted'' to observed properties of the particular realistic 
system. On the other hand, we could use this system to study, in general, the 
dynamical pathways of a simple first order solid state phase transition
involving a structural transition. It is this context that we would like 
to emphasize. Once the equilibrium properties are determined, we look at 
the nucleation dynamics, growth modes and microstructure of the rhombic 
phase growing in the matrix of the parent square lattice\cite{raosen}. 
The effect of defects such as vacancies and dislocations
are automatically incorporated in our microscopic approach. 
In future, we hope to 
obtain atomistically detailed information about the statics and
dynamics of solid-solid interfaces.

\section{Acknowledgement}
Discussions with S. Sastry and G. Baskaran are gratefully acknowledged.
MR thanks the Department of Science and Technology for a Swarnajayanthi
Fellowship.

\section{appendix}
In general, evaluation of three body energies requires sums over all possible 
triplets which for a system of reasonable size is prohibitively expensive.
The particular form for the three body potential used by us is, however, 
special and can be evaluated without keeping track of triplets. 
We illustrate below how this may be done for our system\cite{poten,biswas} 
and derive an explicit expression for the energy.

The three body part of the energy (see Eq. (1) ) is given by,
\begin{eqnarray}
E_{3} & = & \frac{1}{6}\sum_{i \neq j \neq k} \Psi_3(
{\bf r}_{ij},{\bf r}_{jk},{\bf r}_{ki}) \nonumber \\
& = & \frac{1}{2}\sum_{i \neq j \neq k} \frac{V_3}{4}f_{ij}\sin^2 (4 \theta_i) f_{ik} \nonumber \\
& = & \sum_{i \neq j \neq k} 2 (\sin^2\theta_i \cos^2\theta_i - 4 \sin^4\theta_i\cos^4\theta_i) f_{ij}f_{ik}
\end{eqnarray} 

Now define $\tilde{x}_{ij} = x_{ij}/r_{ij}$ and $\tilde{y}_{ij} = 
y_{ij}/r_{ij}$ so that $\sin \theta_i = \tilde{x}_{ik}\tilde{y}_{ij} - 
\tilde{x}_{ij}\tilde{y}_{ik}$ and $\cos \theta_i = \tilde{x}_{ik}
\tilde{y}_{ij} + \tilde{x}_{ij}\tilde{y}_{ik}$. Using the above definitions and the 
quantities,
\begin{eqnarray} 
g_{ij}(1) & = & \tilde{x}^2_{ij}\tilde{y}^2_{ij} f_{ij} \nonumber \\
g_{ij}(2) & = & \tilde{x}^2_{ij}\tilde{y}^2_{ij}(\tilde{x}^2_{ij}-\tilde{y}^2_{ij}) f_{ij} \nonumber \\
g_{ij}(3) & = & \tilde{x}^4_{ij}\tilde{y}^4_{ij} f_{ij} \nonumber \\
g_{ij}(4) & = & \tilde{x}^2_{ij}\tilde{y}^2_{ij}(\tilde{x}^2_{ij}-\tilde{y}^2_{ij})^2 f_{ij} \nonumber \\
g_{ij}(5) & = & \tilde{x}^3_{ij}\tilde{y}^3_{ij}(\tilde{x}^2_{ij}-\tilde{y}^2_{ij}) f_{ij}
\end{eqnarray}
We get, $E_3 = V_3 \sum_i S_i$ with, 
\begin{eqnarray}
S_i & = & 4 \,[G_i(1)F_i-4\,G_i(1)^2-G_i(2)^2]\,-\,16\times\nonumber \\
    &   & \lbrace G_i(3)F_i + 32\,G_i(3)^2 + 2\,G_i(4)^2 + G_i(1)^2\,-\, \nonumber \\ 
    &   &16 \,G_i(3) G_i(1) - 4\,G_i(5)G_i(2) + 16\,G_i(5)^2 \rbrace
\end{eqnarray}
and $G_i(n) = \sum_{j \neq i} g_{ij}(n)$ and $F_i = \sum_{j \neq i} f_{ij}$. 
The three body forces can be got by taking derivatives of $E_3$ which can be 
cast into a similar form.


\newpage
\begin{figure}[]
\begin{picture}(70,70)
\put(0,0) {\epsfig{figure=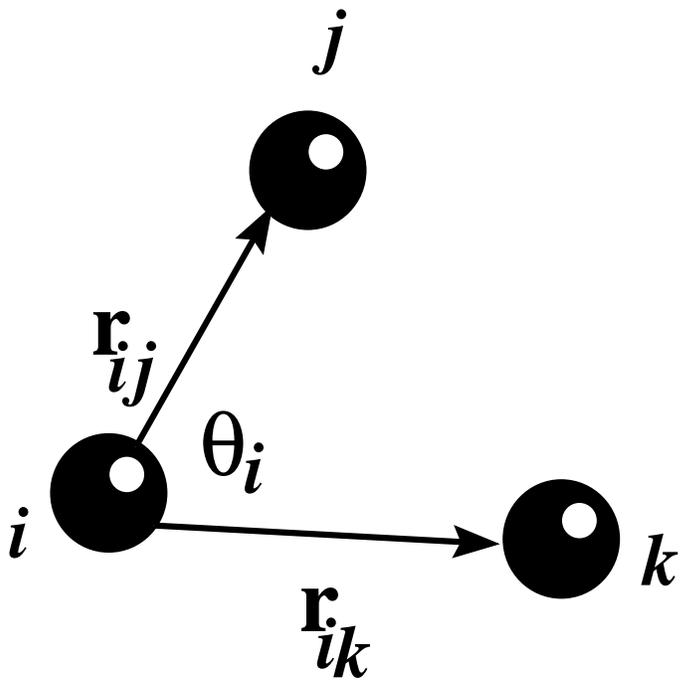,width=90mm,height=90mm}}
\end{picture}
\vskip 2 cm
\caption[]
{\label{angles}
Definition of angles and distances used in the 3-body potential.
}
\end{figure}   

\newpage

\newpage

\newpage

\newpage

\newpage

\newpage

\newpage

\newpage

\newpage

\end{document}